\documentclass[reprint,amsmath,amssymb,aps,prb,floatfix,a4paper]{revtex4-1}

\usepackage{graphicx}
\usepackage{dcolumn}
\usepackage{bm}
\usepackage{color}
\usepackage[ugly]{nicefrac}
\usepackage[latin1]{inputenc}
\usepackage{subfig}
\usepackage{SIunits}

\newcommand{\mrm}[1]{$\mathrm{{#1}}$}
\newcommand{\D}[1]{$\mathrm{\Delta_{{#1}}}$}
\newcommand{\etal}{$\textit{et\ al.\ }$}

\newcommand{\EF}{$\mathrm{E_F}$ }

\begin{document}
\title{Surface spin polarization of the non-stoichiometric Heusler compound $\mathrm{Co_{2}Mn_{\alpha}Si}$.}
\author{Jan-Peter W\"ustenberg}
\email{jpwuest@physik.uni-kl.de}
\author{Roman Fetzer}
\author{Martin Aeschlimann}
\author{Mirko Cinchetti}
\affiliation{Department of Physics and Research Center OPTIMAS, University of Kaiserslautern, Erwin-Schr\"odingerstr.\,46, 67663 Kaiserslautern, Germany}
\author{Jan Min\'ar}
\author{J\"urgen Braun}
\author{Hubert Ebert}
\affiliation{Department Chemie, Ludwig-Maximilians-Universit\"at M\"unchen, Butenandtstra{\ss}e\,11, 81377 M\"unchen, Germany}
\author{Takayuki Ishikawa}
\author{Tetsuya Uemura}
\author{Masafumi Yamamoto}
\affiliation{Division of Electronics for Informatics, Hokkaido University, Kita 14 Nishi 9, Sapporo 060-0814, Japan}

\date{\today}
\begin{abstract}
Using a combined approach of spin-resolved photoemission spectroscopy, band structure and photoemission calculations we investigate the influence of bulk defects and surface states on the spin polarization of \mrm{Co_2Mn_{\alpha}Si} thin films with bulk $L2_1$ order. For Mn-poor alloys the spin polarization at \mrm{E_F} is negative due to the presence of Co$_\mathrm{Mn}$ antisite and minority surface state contributions. We show that in Mn-rich alloys, bulk Co$_\mathrm{Mn}$ antisites are suppressed which leads to a positive spin polarization at the Fermi energy, and the influence of minority surface states on the photoelectron spin polarization is reduced. 
\end{abstract}
\maketitle
\section{Introduction}
\label{sec:intro}
Heusler compounds are intermetallic compounds with the composition \mrm{X_2YZ} that crystallizing in the $L2_1$ structure. X and Y atoms are transition metals and Z is a main group element. Due to a high Curie temperature and predicted full conduction electron spin polarization, some members of the familiy of Heusler compounds are expected to improve significantly the performance of magnetic tunneling junctions (MTJs) based on the tunneling magnetoresistance effect~\cite{Balke08}.

Among the class of ferromagnetic Heusler compounds, $\mathrm{Co_2MnSi}$ (CMS) has been most successfully applied in MTJs~\cite{Ishikawa09a,Sakuraba10a}. It is predicted to be a half metallic ferromagnet with a large magnetic moment of \unit{5}{\mu_B} per formula unit and a high Curie temperature of \unit{985}{\kelvin}~\cite{Ishida95,Chadov09,Chioncel08,Webster71}. All calculations predict a nearly half metallic behavior of bulk single crystals owing to a gap in the minority density of states around the Fermi energy, at least at low temperature. In contrast to the large bulk band gaps predicted by density functional theory (DFT) calculations (cf.\ Ref.~\onlinecite{Chadov09}) the experimental width of this band gap in thin film structures has  been found to be about \unit{0.4}{eV} using tunneling spectroscopy at interfaces with different barrier materials~\cite{Sakuraba06,Ishikawa09}. 

In order to maintain a high electron spin polarization even at room temperature the Fermi energy should lie preferentially in the middle of the half metallic gap. In this aspect, experiments and theory disagree considerably. While DFT calculations find the Fermi energy in CMS to be situated close to the minority valence band, MTJs using MgO and AlO$_x$ barriers indicate a Fermi energy around midgap~\cite{Ishikawa09} and \unit{10}{meV} below the conduction band~\cite{Sakuraba06}, respectively. 

So far, experimental spin polarization values have not reached the ideal limit of \unit{100}{\%} at the Fermi energy. In most publications, an effective spin polarization is inferred from the tunnel magnetoresistance ratio (TMR) of MTJ by making use of the Julli\`{e}re formula $TMR=2P_\mathrm{J1}P_\mathrm{J2}/(1-P_\mathrm{J1}P_\mathrm{J2})$, where $P_\mathrm{J1}$ and $P_\mathrm{J2}$ denote the effective spin polarization values for the respective ferromagnet/insulator interface~\cite{Julliere75}. A high tunneling spin polarization of $P_\mathrm{J}=\unit{89}{\%}$ has been inferred using the tunneling magneto-resistance ratio (TMR) of a CMS/AlOx/CMS MTJ at \unit{2}{K}, decreasing to \unit{72}{\%} at room temperature~\cite{Sakuraba06}. Point contact Andreev reflection (PCAR) measurements at T=\unit{4.2}{K} on bulk single crystals revealed $P_\mathrm{PCAR}=\unit{59}{\%}$~\cite{Rajanikanth09}. Using spin-resolved photoelectron spectroscopy (SR-PES) carried out at room temperature, a maximum surface spin polarization value of $P_\mathrm{PES}=\unit{12}{\%}$ was obtained for thin films of $\mathrm{Co_{2}MnSi}$ (100) grown on GaAs(001) using \unit{70}{eV} photons~\cite{Wang05}. For Heusler compounds grown on MgO(100) substrates, \mrm{Co_2FeSi}, \mrm{Co_2MnGa} and \mrm{Co_2Cr_{0.6}Fe_{0.4}Al} were investigated by SR-PES so far~\cite{Schneider06,Cinchetti07,Wustenberg07,Wustenberg09,Hahn11}.   

The strong temperature dependence of the TMR of Heusler based MTJs has been related to a temperature dependent spin polarization in the ferromagnetic electrode, via the Julli\`{e}re formula. Most prominently so called non-quasiparticle states have been invoked~\cite{Irkhin07,Chioncel08}. However, high energy photoemission experiments, applied to verify spin integrated calculated valence band densities of states could not confirm the predicted changes in the peak positions~\cite{Fecher08,Miyamoto09}. Other intrinsic bulk depolarization mechanisms such as spin orbit interaction~\cite{Mavropoulos04a}, magnetic sublattice non-collinearity, fluctuation induced hybridization changes and weakened exchange coupling of the magnetic surface layer have been discussed~\cite{Lezaic06,Skomski07} but still lack experimental verification.

Extrinsic effects such as deviations of the \mrm{L2_1} structure due to disorder and non-stoichiometric composition can lead to additional electronic states in the minority gap that reduce the spin polarization at the Fermi energy. Opposite to various non-critical types of defects, \mrm{Co_{Mn}} antisite defects have been predicted to induce minority defect states at the Fermi energy~\cite{Picozzi04,Galanakis06a,Huelsen09}. Recent results suggest that such antisites are hard to avoid in stoichiometric thin films but can at least partly be compensated by increasing the Mn/Co ratio~\cite{Yamamoto10,Ishikawa09a}. However, at large Mn concentrations the precipitation of non-ferromagnetic \mrm{Mn_3Si} clusters has been postulated~\cite{Huelsen09}. 

Additional spin polarized electronic states may arise at surfaces and interfaces due to hybridization changes that occur at the boundary atomic layers~\cite{Luth01}. Such states are well known for surfaces with a gap in the surface-projected electronic band structure, for example in semiconductors or in the projected bulk band gaps at the low index surfaces of various transition metal surfaces~\cite{Luth01}. Such localized states can contribute significantly to the electronic transport properties of tunneling magnetoresistive devices~\cite{Mavropoulos05} and have been used to explain the I/V characteristics of magnetic tunneling junctions~\cite{Ishikawa09}.

For free $\mathrm{Co_{2}MnSi}$ (CMS)(100) surfaces, ab initio atomistic thermodynamics predict four stable terminations: Mn-Mn, Mn-Si, Si-Si and vacancy-Si. Apart from the Mn-Mn termination, all stable surface configurations have been predicted to reduce the surface spin polarization due to the formation of minority surface state bands that are derived from partly filled Co\mrm{d_{3z^2-r^2}} orbitals and cross the Fermi level within the bulk minority band gap~\cite{Hashemifar05}.

In this report we present a systematic characterization of the free (100) surface of Mn-deficient and Mn-rich CMS films in terms of geometric and spin resolved electronic structure. The low energy electron diffraction (LEED) patterns indicate a well-ordered surface. Using low energy spin-resolved photoelectron spectroscopy (SR-PES), the spin-resolved electronic structure is determined and compared to state of the art photoemission calculations performed for different surface terminations and including dynamic correlation effects. We find that both LEED patterns and spin-resolved spectra corroborate the assumption of a  Mn-Si or vacancy-Si terminated surface. The measured spin polarization at the Fermi energy is substantially reduced by the presence of surface states with predominant minority spin character and antisite defects. In particular we find that for Mn-poor alloys the spin polarization at \EF is negative due to the presence of Co$_\mathrm{Mn}$ antisite and minority surface state contributions. In Mn-rich alloys, the suppression of Co$_\mathrm{Mn}$ antisites leads to a positive spin polarization at the Fermi energy, and the influence of minority surface states on the photoelectron spin polarization is reduced.
\section{Calculational details}
\label{sec:theo}
To fully understand the experimental data, we have performed first-principles calculations using the local-spin-density approximation (LSDA) of density functional theory within the spin-polarized fully relativistic Korringa-Kohn-Rostoker Green's function method (SPR-KKR) \cite{EKM11,Ebe00,SPR-KKR5.4}. For the exchange and correlation potential we applied the Vosko, Wilk, and Nusair parametrization \cite{VWN80}. To account for electronic correlations beyond the LSDA we employed a combined LSDA plus dynamical mean field theory (LSDA+DMFT) scheme, self-consistent in both the self-energy calculation and in the charge density calculation, as implemented within the relativistic SPR-KKR formalism \cite{Min11,MCP+05}. As a DMFT-solver the relativistic version of the so-called Spin-Polarized T-Matrix Plus Fluctuation Exchange (SPTF) approximation \cite{KL02,PKL06} was used. In contrast to most other LSDA+DMFT implementations, within the SPR-KKR scheme the complex and energy-dependent self-energy $\Sigma_{DMFT}$ is implemented as an additional energy-dependent potential to the radial Dirac equation which is solved in order to calculate the new Green's function. This procedure is repeated until self-consistency in both the self-energy and the charge density is achieved. The double counting problem (separation of the Hubbard Hamiltonian from the LSDA one) was considered within the usual around atomic limit (AAL). This scheme was successfully used before in describing magnetic properties of CMS \cite{Chadov09}. An appealing feature of the multiple scattering formalism is the possibility to calculate substitutionally disordered materials within the coherent potential approximation (CPA). The CPA is considered to be the best theory among the so-called single-site (local) alloy theories that assume complete random disorder and ignore short-range order.  A combination of the CPA and LSDA+DMFT within the SPR-KKR method has been used recently \cite{MCP+05,CMK+08,SMME08}.

The self-energy within the DMFT is parametrized by the average screened Coulomb interaction $U$ and the Hund exchange interaction $J$. The $J$ parameter can be calculated directly within the LSDA and is approximately the same for all 3$d$ elements; we used $J_{\text{Mn,Co}}$=0.9~eV for the Mn and Co atoms throughout our work. The parameter $U$ is strongly affected by the metallic screening and it is estimated for the 3$d$ metals between 1-3~eV. We used $U_{\text{Mn,Co}}$=2.3~eV for the Mn and Co atoms \cite{Chadov09}. 
DMFT calculations have been performed for T=\unit{400}{K} and we used $4096$ Matsubara poles to calculate the corresponding SPTF self-energy. The effective potentials were treated within the atomic sphere approximation (ASA). 

As a first step of our theoretical investigations, we performed LSDA+DMFT self-consistent electronic structure calculations for ordered bulk \mrm{Co_2MnSi} and CPA calculations for three different disordered compositions (\mrm{Co_2(Mn_{0.69}Vacancy_{0.31})Si}, \mrm{Co_2(Co_{0.168}Mn_{0.748}Si_{0.084})Si} (corresponding to a Mn-poor compound of nominal \mrm{Co_2Mn_{0.69}Si} composition) and \mrm{(Co_{1.909}Mn_{0.091})Mn_{1.0}(Si_{0.955}Mn_{0.045})} (corresponding to a Mn-rich compound of nominal \mrm{Co_2Mn_{1.19}Si} composition). For all systems we used the L2$_1$ structure with the experimental lattice constant (a$_{lat}$=5.654$\AA$). In addition, to study surface effects for free $\mathrm{Co_{2}MnSi}$(100) surface, semi infinite screened KKR calculations \cite{EKM11} (i.e.\ without using artificial slab geometry) have been performed for four stable terminations: Mn-Mn, Mn-Si, Si-Si and vacancy-Si. For the multipole expansion of the Green's function, an angular momentum cutoff of $l_{\text{max}}=3$ was used. The integration in the $\bm{k}$\ space was performed by the special points method using 1600 $\bm{k}$ points in the irreducible wedge. 

As a second step, actual valence band photoemission spectra were calculated using a recent implementation of the fully relativistic CPA formalism within the LSDA+DMFT method in the framework of the one step model of photoemission, which implicitly includes all matrix elements and surface effects \cite{BMM+10,MBME11} and therefore allows for a direct comparison to the corresponding experimental data.

\section{Experiment}
\label{sec:exp}
We have investigated two different \mrm{Co_2Mn_{\alpha}Si(100)} surfaces. Both films showed \mrm{L2_1} bulk order in the XRD patterns. Sample CMS069 with $\alpha$ = 0.69 exhibited a nominal bulk \mrm{Co_2Mn_{0.69}Si_{1.01}} composition, while sample CMS119 with $\alpha$ = 1.19 exhibited \mrm{Co_2Mn_{1.19}Si_{0.88}} composition. The layer structure of the samples was as follows: (from the substrate side) MgO buffer(\unit{10}{nm})/CMS($\alpha$)/MgO barrier(\unit{2}{nm}), grown on a MgO(100) substrate. The CMS thickness was \unit{50}{nm} for CMS069 or \unit{30}{nm} for CMS119. Each layer was successively deposited in an ultrahigh vacuum chamber (with a base pressure of about \unit{6\times 10^{-10}}{mbar}). The CMS layer was deposited at room temperature at RT using magnetron sputtering and subsequently annealed in situ at \unit{600}{ºC}. The MgO(100) barrier was deposited at RT by electron beam evaporation. Each layer in the above layer structure was grown epitaxially~\cite{Yamamoto10}. Before LEED observations or spin-resolved photoemission measurements, samples were annealed to \unit{500}{ºC} in situ in an ultrahigh vacuum chamber for observations and measurements. All measurements were carried out at room temperature.

LEED patterns of the sample surfaces were obtained by means of an Omicron 3-grid SpectaLEED system after a cleaning procedure, consisting of the removal of the MgO barrier by \unit{500}{eV} Ar$^+$ sputtering and subsequent annealing of the CMS samples to \unit{500}{ºC}. From the ratio of spot width (Lorentz FWHM) and reciprocal lattice vector an instrumental transfer width~\cite{Ertl85} of at least \unit{12}{nm} was determined at a clean Cu(100) reference surface and at \mrm{E=\unit{80}{eV}}. The chemical composition was investigated using an Auger system based on a cylindrical mirror energy analyzer, manufactured by Omicron NanoTechnology. Relative composition values were obtained using the method described in Ref.~\onlinecite{Davis87}.

The photoemission spectra were obtained using the linearly polarized 4th harmonic (photon energy \unit{5.9}{eV}) of a \unit{100}{fs} Ti:Sapphire oscillator (Spectra Physics Tsunami). Using a phase retarding plate the light polarization could be switched from s polarization (electrical field vector perpendicular to the plane of incidence) to p polarization (electrical field vector within the plane of incidence). The photon angle of incidence onto the samples was \unit{45}{°}. The photoelectron spectra were taken in normal emission geometry, integrating along the $\mathrm{\Gamma X}$ momentum direction. A biasing voltage of \unit{-4}{V} was applied between sample and detector in order to increase the effective parallel momentum integration window. Due to the large unit cell size of CMS and the large angular acceptance angle of the analyzer of $\pm$\unit{15}{°} along the sample {[}110{]} direction (\mrm{\Gamma K}=\mrm{\bar{\Gamma}\bar{X}} momentum direction), the photoemission spectrum integrates electron momenta over almost \unit{60}{\%} of the surface Brillouin zone of \mrm{Co_2MnSi} in $\Gamma K$ direction.  

Spin polarized photoemission spectra were recorded by means of a commercial \unit{90}{°} cylindrical sector energy analyzer (Focus CSA 300), equipped with an additional spin detector based on spin-polarized LEED at a W(100) crystal (Focus SPLEED). The achieved effective energy resolution is \unit{210}{meV} full width at half maximum as determined from a simulation of the work function cutoff of the spectra.

Due to the detection geometry, the electron spin polarization (ESP) along the CMS~[011] direction (in-plane) and along the surface normal (out-of-plane) can be determined. A Sherman factor of S=$0.2$ was used to determine the spin polarization from the measured intensity asymmetry of opposite diffraction spots. Apart from statistical errors, a systematic relative spin polarization error due to uncertainties in the determination of the Sherman factor (S=0.2\ldots 0.25) can not be excluded~\cite{Kirschner85}. Minority and majority spin spectra were calculated using the formula  $I_{\uparrow\downarrow}=\bar{I}\left(1\pm P\right)$ using the experimental ESP $P$ and the averaged count rate $\bar{I}$ of corresponding opposite channeltrons. For the spin-resolved measurements the CMS film was remanently magnetized along the in-plane [011] direction by applying an external in-plane magnetic field of \mrm{\mu_{0}H=\unit{15}{mT}}, which is sufficient to saturate the magnetization of the investigated samples. Detector related asymmetries are canceled out by taking separate measurements with reversed sample magnetization. With this setup an in-plane spin polarization of \unit{30}{\%} at the Fermi energy was achieved on a \unit{200}{nm} thick film of polycrystalline $\mathrm{Co_{70}Fe_{30}}$. The out-of-plane spin component vanished within the experimental error.
\section{Surface characterization}
\label{sec:augerleed}
Before investigating the surface electronic structure of \mrm{Co_2MnSi}(100) a clean and well defined sample surface was prepared. This was accomplished by sputter cleaning using \unit{500}{eV} Ar$^+$ ions and subsequent heating of the sample. The surface quality was confirmed by Auger spectroscopy and by LEED pattern analysis.

\begin{figure}
\centering
\includegraphics[width=\linewidth]{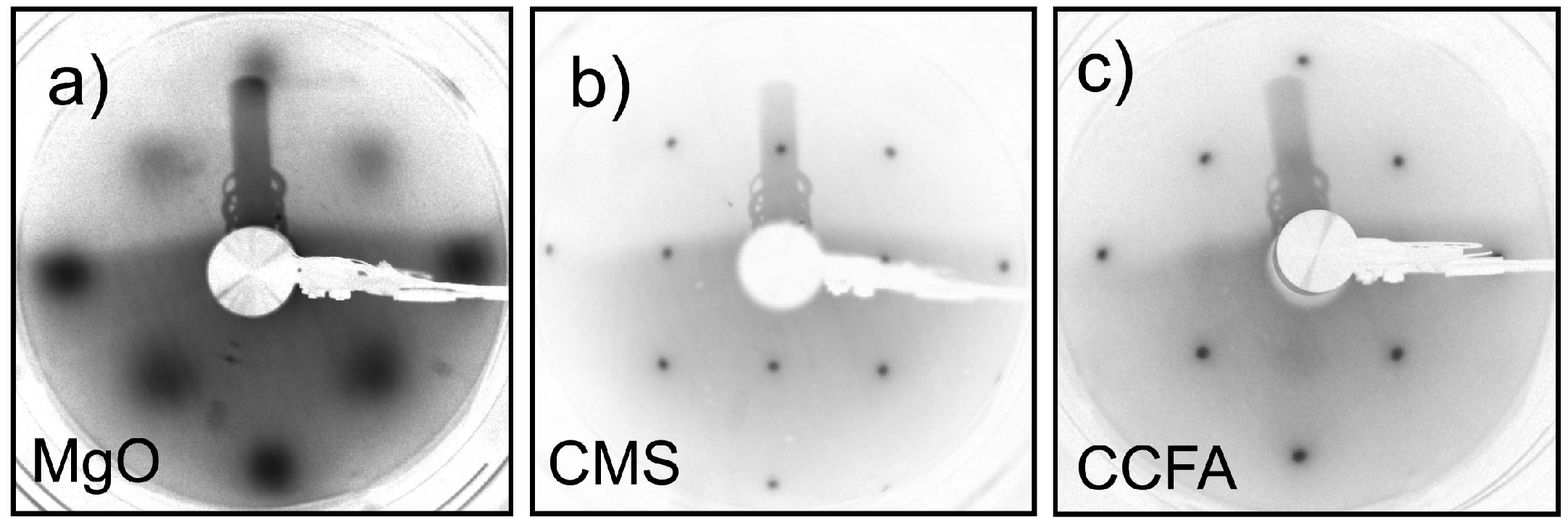}
\caption{LEED patterns of a) a MgO barrier (E=82\,eV), b) sample CMS069 (E=57\,eV) and c) \mrm{Co_2Cr_{0.6}Fe_{0.4}Al} (E=74\,eV).}
\label{fig:pattern}
\end{figure}

In Figure~\ref{fig:pattern} the LEED patterns of an MgO tunneling barrier (B1 structure), sample CMS069 and a \mrm{Co_2Cr_{0.6}Fe_{0.4}Al} (B2 structure~\cite{Herbort09,Wustenberg09}) are shown. The pattern of CMS069 and CMS119 (not schown) are identical. Sharp spots indicate a well ordered CMS surface. Comparing with the MgO(100) surface pattern, we observe the compressed and rotated spot pattern expected from the larger unit cell CMS and the resulting \unit{45}{°} growth in order to minimize lattice misfit. Moreover, an (effectively) monoatomic surface termination as shown in the \mrm{Co_2Cr_{0.6}Fe_{0.4}Al} pattern can be excluded by the presence of the (11) edge spot, indicating a square two-atomic surface lattice. Thus we conclude that the (100) surface of the samples CMS069 and CMS119 is indeed terminated by a Mn-Si (or vacancy-Si) layer as reported in earlier publications on stoichiometric CMS~\cite{Miyajima09,Wang05,Hashemifar05}.
\begin{figure}
\centering
\includegraphics[width=\linewidth]{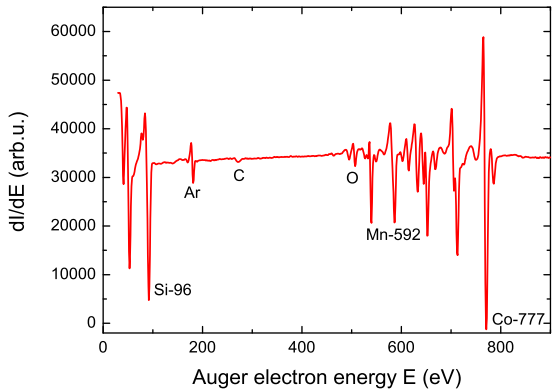}
\caption{Auger spectrum of the free CMS069\,(100) surface, obtained at \unit{3}{kV} primary electron energy}
\label{fig:Auger}
\end{figure}

In order to determine the chemical composition of the CMS069 surface, Auger spectroscopy has been carried out. The obtained spectrum of sample CMS069 is displayed in Fig.~\ref{fig:Auger}. It clearly shows the signatures of the constituting elements, along with some residual argon from the sputtering process, carbon (\unit{1}{\%}) and oxygen (\unit{3}{\%}) which could not be removed without leaving the optimum annealing temperature range. Using the peak-to-peak amplitudes of the differential spectrum and the sensitivity factors given in Ref.~\onlinecite{Briggs03} the composition of the surface was obtained. The analysis of the differentiated Auger spectrum yields a nominal surface composition of \mrm{Co_2Mn_{0.85}Si_{0.93}}, thus a slightly higher Mn content than the thin film value determined by ICP analysis~\cite{Yamamoto10}. For CMS119, we obtain a nominal surface composition of \mrm{Co_2Mn_{1.19}Si_{1.35}}. 
\section{Electronic structure}
\subsection{SR-PES spectra}
\label{sec:specs}

\begin{figure}
\centering
\includegraphics[width=\linewidth]{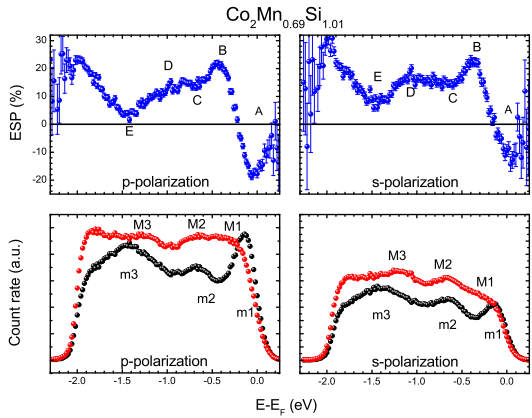}
\caption{Spin resolved photoemission spectra of the free CMS069\,(100) surface for p-polarized excitation (left) and s-polarized excitation (right). Upper panel: electron spin polarization (ESP), lower panel: majority and minority count rates. ESP features are marked by A\ldots E, Majority (minority) features are labelled M1\ldots M3 (m1\ldots m3).}
\label{fig:CMS069}
\end{figure}

\begin{figure}
\centering
\includegraphics[width=\linewidth]{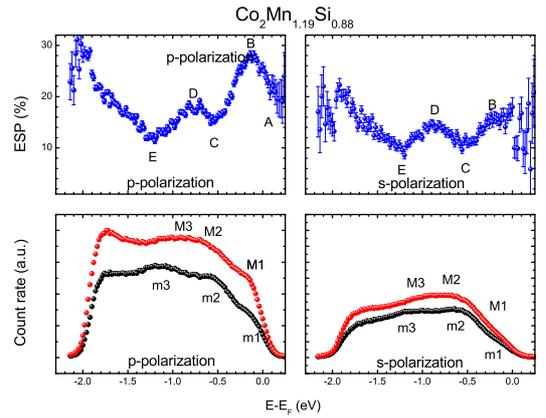}
\caption{Spin resolved photoemission spectra of the free CMS119\,(100) surface for p-polarized excitation (left) and s-polarized excitation (right). Upper panel: electron spin polarization (ESP), lower panel: majority and minority count rates. ESP features are marked by A\ldots E, Majority (minority) features are labelled M1\ldots M3 (m1\ldots m3).}
\label{fig:CMS119}
\end{figure}

In Figures \ref{fig:CMS069} and \ref{fig:CMS119} we show the spin resolved photoemission spectra obtained respectively from the freshly prepared CMS069 and CMS119 surfaces, using p and s polarized laser pulses for excitation. As expected from the chemical and  structural similarity, both samples show similar spectral features, labeled with M1\ldots M3 (m1\ldots m3) in the majority (minority) spectra. However, the resulting electron spin polarization (ESP) shows pronounced differences in its features, labeled (A)\ldots (E). 

The most striking differences between CMS060 and CMS119 are observed at feature (A) in the energy region around the surface Fermi level. In CMS069, a negative ESP is observed for both s and p polarized excitation, while in CMS119, the ESP in (A) is positive for both light polarizations, but still far from \unit{100}{\%}. 

Towards lower energies the ESP recovers to values between $+0.2$ and $+0.3$ (B). Continuing further downwards in energy we observe a slight minimum (C), followed by a maximum (D). The latter two features are more pronounced in CMS119 as compared to CMS069. A minimum (E) in the ESP is visible for all samples. However, its energy is shifted by \unit{0.2}{eV} towards \EF in CMS119 (compare Tab.~\ref{tab:SP}). We do not observe any peak shifts due to a change of the light polarization.
\begin{table}[htbp]
  \centering
  
    \begin{tabular}{l|rrrrr}
    & \multicolumn{1}{c}{{A}} & \multicolumn{1}{c}{{B}} & \multicolumn{1}{c}{{C}} & \multicolumn{1}{c}{{D}} & \multicolumn{1}{c}{{E}} \\
    \hline \hline
	{CMS069-p} & -0.18 & -0.44 & -0.68 & -0.96 & -1.44 \\
    {CMS069-s} & 0.00     & -0.38 & -0.64 & -1.08 & -1.44 \\
    \hline
    {CMS119-p} & 0.00     & -0.14 & -0.54 & -0.77 & -1.24 \\
    {CMS119-s} & \multicolumn{1}{c}{--} & 0.00     & -0.50  & -0.88 & -1.24 \\
    \end{tabular}%
  \caption{Energetic position of the described spin polarization features given in eV with respect to \EF. }
  \label{tab:SP}%
\end{table}%

In order to identify spectral features responsible for the observed ESP we have plotted the majority and minority spectra in the lower panels of Figs. \ref{fig:CMS069} and \ref{fig:CMS119}. The energetic positions of the most dominant spectral features are summarized in Tab.~\ref{tab:Mm}. First we note that the spectral peak positions do not much depend on the polarization direction of the incident photons.  
\begin{table}[htbp]
  \centering
  \begin{tabular}{l|rrrrrr}
     & \multicolumn{1}{c}{m1} & \multicolumn{1}{c}{m2} & \multicolumn{1}{c}{m3} & \multicolumn{1}{c}{M1} & \multicolumn{1}{c}{M2} & \multicolumn{1}{c}{M3} \\
    \hline \hline
    {CMS069-p} & -0.14 & -0.68 & -1.44 & -0.20 & -0.64 & -1.28 \\
    {CMS069-s} & -0.16 & -0.66 & -1.42 & -0.20 & -0.68 & -1.20  \\
    \hline
    {CMS119-p} & -0.18 & -0.60 & -1.16 & -0.22 & -0.70 & -0.92  \\
    {CMS119-s} & -0.18 & -0.64 & -1.16 & \multicolumn{1}{c}{--} & -0.58 & -0.92  \\
    \end{tabular}%
  \caption{Peak energies within the minority (m) and majority (M) spectra given in eV with respect to \EF. }  
  \label{tab:Mm}%
\end{table}%

As we can clearly see in Fig.~\ref{fig:CMS069}, the negative ESP (A) of CMS069 at the Fermi energy is related to a strong peak m1 in the minority electron spectrum, with a higher contribution to the spectrum for p polarized excitation than for s polarized excitation. The recovery of the ESP (B) is reflected in a local minimum of minority intensity following the minority peak at the Fermi energy. 

In sample CMS119,  the spectral features around \EF are less pronounced than for CMS069 but nevertheless show a clear polarization dependence (see Fig.~\ref{fig:CMS119}). In contrast to CMS069, the ESP at \EF is dominated by the majority spin channel. For p polarized light, the majority DOS is clearly metallic, leading to the strong maximum (B). For s polarized excitation, the maximum (B) is less pronounced, which is reflected in a less well defined Fermi level cutoff for majority electrons. The minority DOS around \EF is strongly reduced as compared to CMS069. The ESP minimum (C) is formed by the features m2 and M2 in both minority and majority spectra, with similar peak positions in both samples. The small ESP increase at the low energy end of the plateau is caused by the broad majority peak M3. The latter and the minority peak m3 in CMS119 are both shifted by almost the same amount of \unit{0.2}{eV} towards \EF with respect to CMS069, causing the observed shift of the ESP minimum (E).

\subsection{Initial state symmetries}
\label{sec:symm}
\begin{figure}
\centering
\includegraphics[width=\linewidth]{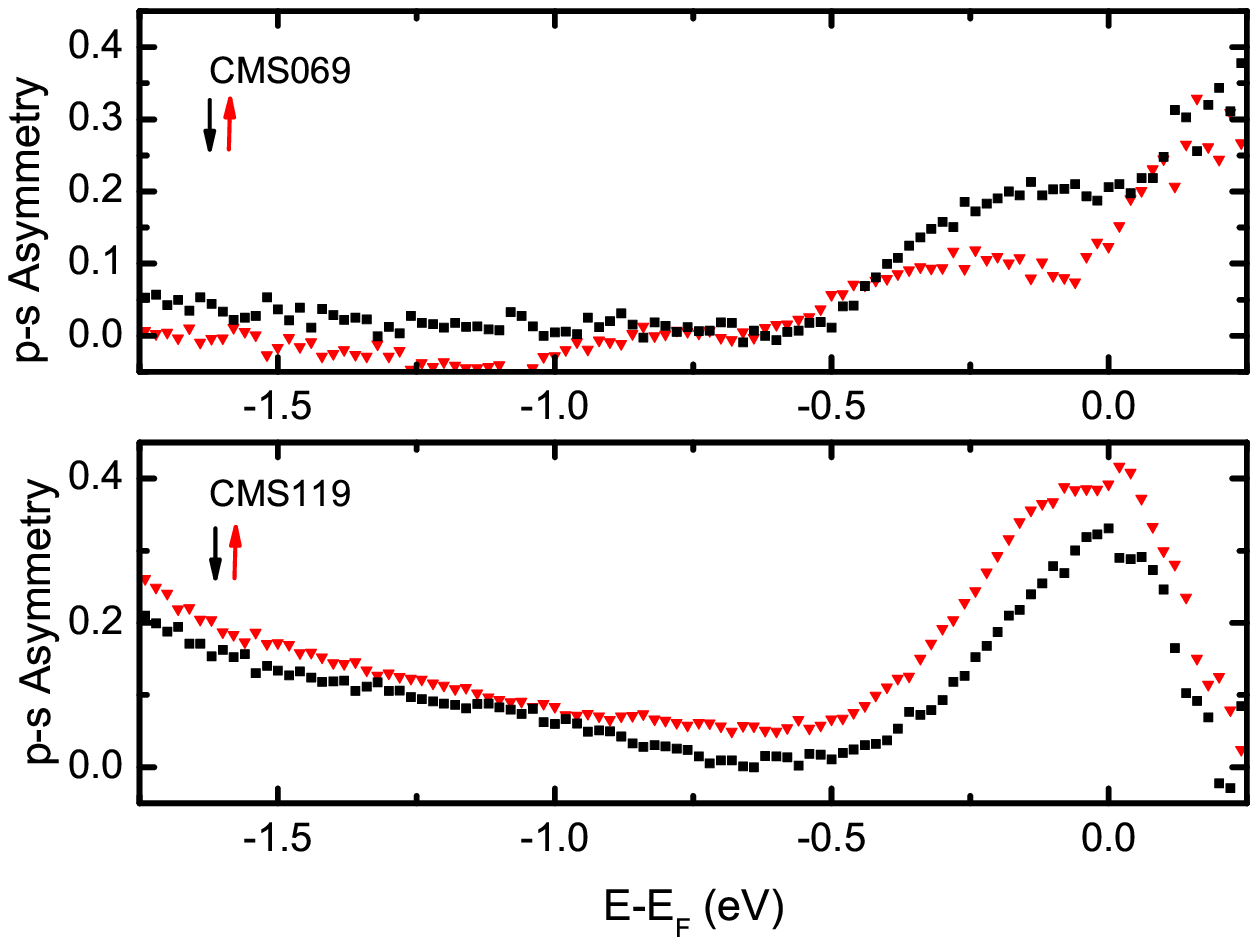}
\caption{Calculated experimental asymmetry spectra for p and s polarized excitation in CMS069 (upper panel) and CMS119 (lower panel). Majority (minority) spin spectra are marked with red triangles (black squares). The spin resolved spectra of each sample and polarization were normalized on the respective minority m2 peak.}
\label{fig:Asym_CMS}
\end{figure}

In this section we identify the symmetry properties of the initial states, giving rise to the intensity differences observed for different light polarization directions (p and s). To this end, we have calculated the ps-asymmetry spectra $A_{ps}=\frac{I_p-I_s}{I_p+I_s}$ for majority and minority electrons of each sample (see Fig.~\ref{fig:Asym_CMS}). Here, $I_p$ and $I_s$ are the intensities of the respective photon polarization. To compensate for the larger photoelectron yield when using p-polarized light, the intensities of the spin resolved spectra have been normalized to the intensity of the minority peak m2 for p and s polarized excitation, respectively. If both polarization directions would address only states with the same symmetry, the ps-asymmetry would be a constant, independent of the normalization. 

The observed ps-asymmetry is caused by matrix element effects due to the different spatial symmetry properties of the involved initial states, and indicates the reflection symmetry of the initial states with respect to a mirror plane normal to the sample surface. While s-polarized light has only components along the in-plane \mrm{[01\bar{1}]} direction and thus only excites states with \mrm{\Delta_5} symmetry, p-polarized light additionally excites states of \mrm{\Delta_1} symmetry in our experimental geometry, due to its field component along the normal [100] direction~\cite{Hermanson77}. The variation of the ps-asymmetry spectrum thus can be used as a measure for the contribution of states compatible with \D{1} symmetry, be it a bulk, defect or a surface related transition. The background is formed by a common contribution originating from \D{5} initial states that contribute to the spectrum of both light polarization directions.

The experimental ps-asymmetry of majority (M) and minority (m) electron spectra for CMS069 and CMS119 are displayed in Fig.~\ref{fig:Asym_CMS}. 
We observe a strong ps-asymmetry maximum around the Fermi energy \EF in both samples, for both majority and minority electrons. This asymmetry peak extends down to \unit{-0.6}{eV}, with its maximum situated close to the features M1 and m1.  We observe a clear substructure in CMS069, which can be related to the shift between the center of gravity of the peak M1 and the sharp peak m1. A relative minimum is reached close to the position of the features m2 and M2. Towards lower energies, the \D{1} contribution stays stable (CMS069) or rises slightly (CMS119). In contrast to CMS069, the \D{1} contribution in CMS119 is stronger in the majority spectrum. This could explain the higher TMR ratios in MgO based MTJs for Mn-rich CMS electrodes, where the transmission of wave functions with \D{1} symmetry is favored~\cite{Miura07}. 

\subsection{Resonant bulk transitions}
\label{sec:bulk}
\begin{figure}
\centering
\subfloat[Majority]{\includegraphics[width=0.6\linewidth]{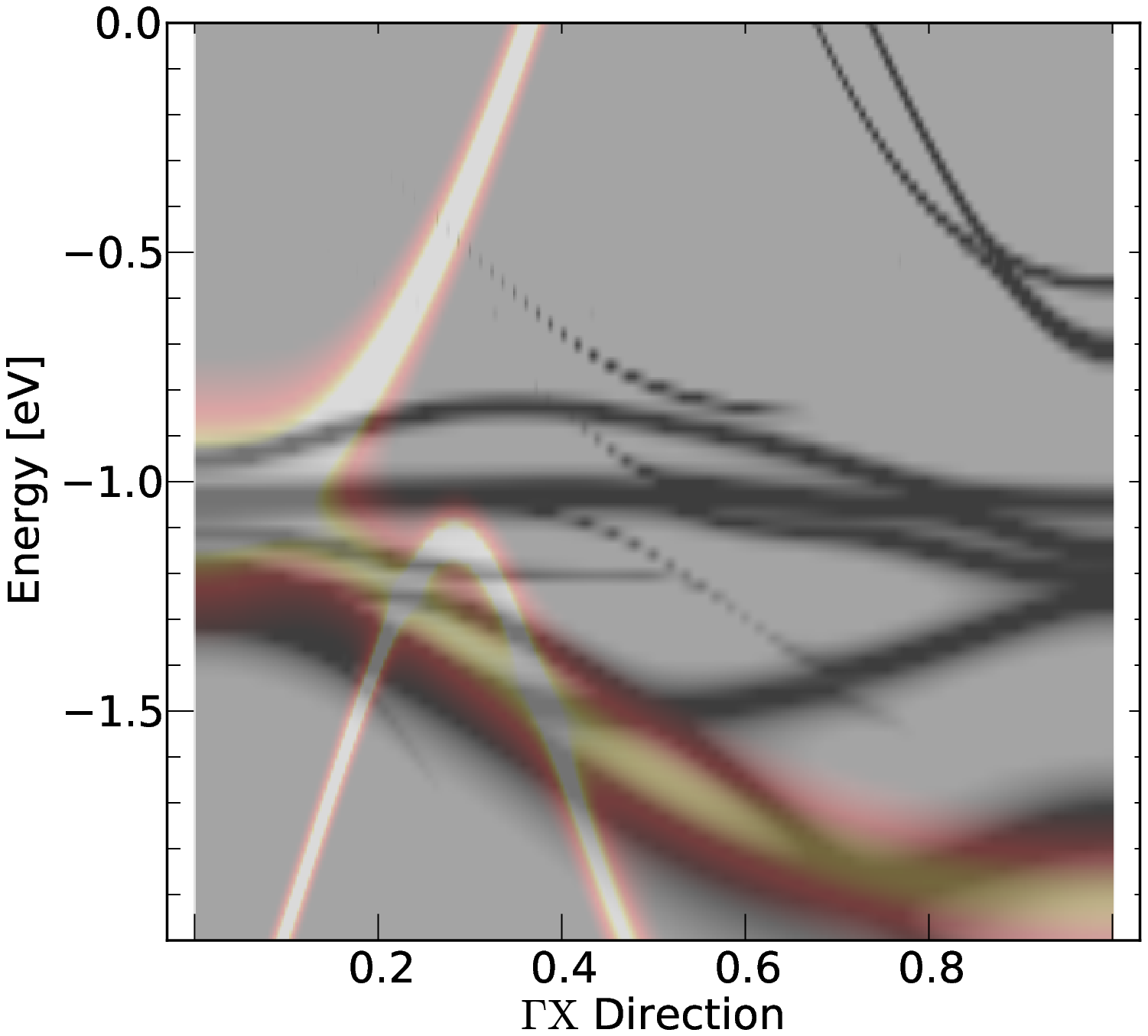}}\\
\subfloat[Minority]{\includegraphics[width=0.6\linewidth]{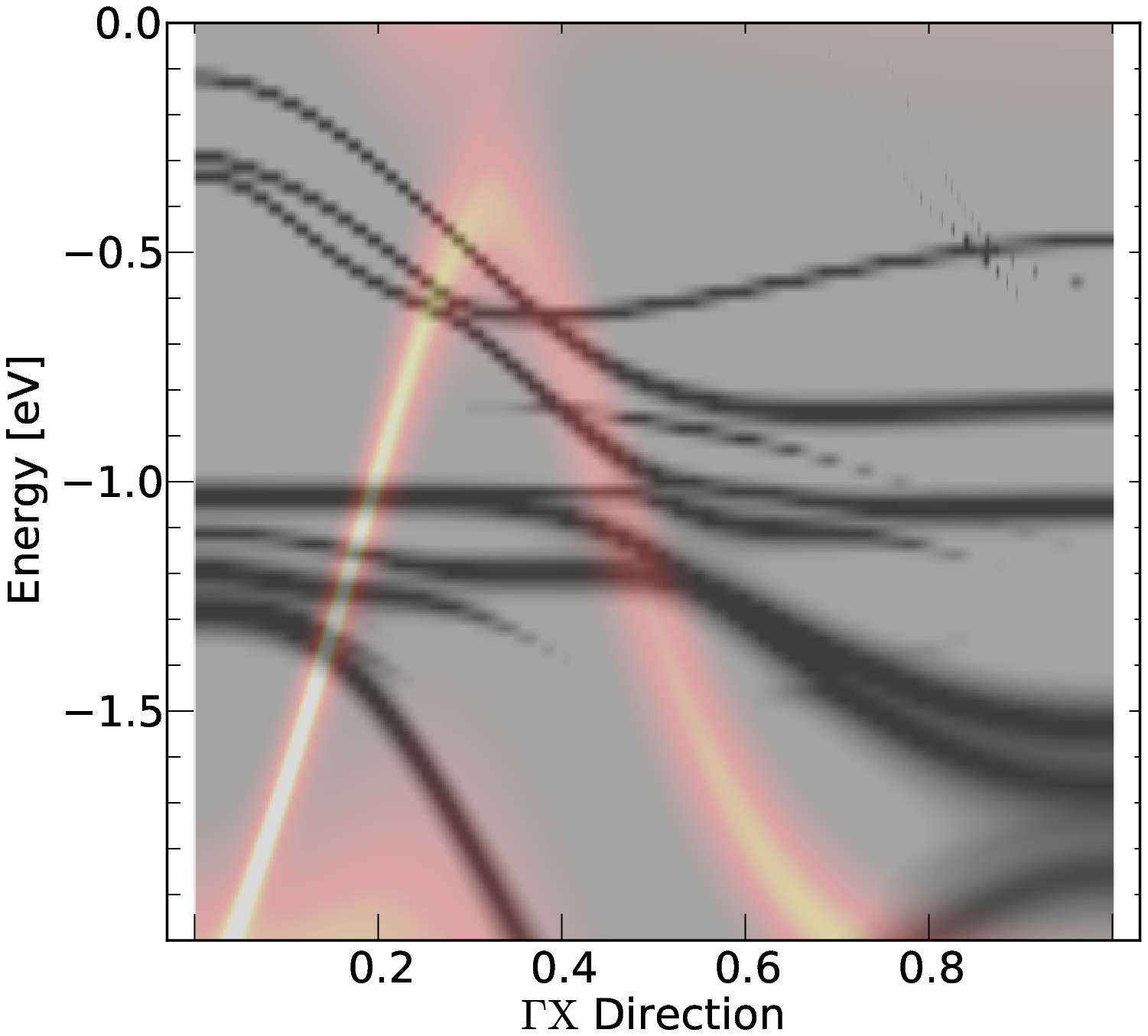}}
\caption{Possible bulk transitions for majority and minority electrons along the \mrm{\Gamma X} momentum direction of CMS as calculated using LSDA+DMFT. Occupied bulk states (unoccupied final states) are shown as black (orange) lines. Crossing points correspond to energy and momentum allowed transitions at the photon energy of $\hbar\omega$=\unit{5.9}{eV}.}
\label{fig:CMS_Trans}
\end{figure}

In this section we identify the possible contribution of bulk electronic transitions and final state effects to the experimental spectra in section~\ref{sec:exp}. For this purpose we calculated the bulk spectral functions of stoichiometric CMS along the surface normal using LSDA+DMFT. In Fig.~\ref{fig:CMS_Trans} we show an overlay of the ideal $\mathrm{L2_{1}}$ occupied bands (black) and a replica (orange) of the unoccupied final states, where the latter are shifted downwards by the experimental photon energy of $\mathrm{\hbar\omega}$=5.9\,eV. In this way, energy ($\Delta E=\hbar \omega)$ and momentum-allowed ($\Delta \mathbf{k}=0$) bulk resonant interband transitions appear as crossing points between occupied and unoccupied bands and can directly be compared to the energies of measured spectral features. Matrix element effects such as symmetry selection rules are not considered in this picture. Since the momentum conservation is broken along the surface normal, all transitions along the \mrm{\Gamma X} direction are expected to contribute to the measured photoelectron spectrum.

First we note that there are no allowed resonant bulk majority transitions in the vicinity of the Fermi energy. Resonant bulk transitions in the majority spin direction can only be found below \mrm{E-E_F=-0.9\,eV}, They involve localized orbitals as initial states and most probably constitute the main contribution to the broad majority M3 peak. Above this energy, we find dispersive bulk majority electron states with \D{1} and \D{5} symmetry~\cite{Miura07}, with their band bottom situated \unit{0.6}{eV} and \unit{0.7}{eV} below the Fermi energy, respectively. These states might be the origin of surface resonances that form the complex peak consisting of M1 and M2 in the majority spectra, and thus might contribute significantly to the measured ESP in the vicinity of \EF.

For minority electrons, possible resonant bulk transitions can be found starting 0.5~eV below the bulk valence band maximum. These transitions stem from the dispersive minority valence band and contribute to the minority peak m2 in the spectra. Between -1.0~eV and -1.4~eV a second, more localized set of resonant transitions appears which we assign to the positions of peak m3 in the minority spectra. Due to their flat dispersion, these states should be robust against final state effects. 

In the vicinity of the Fermi energy, only non-resonant optical minority electron transitions are possible, and no intensity should be expected from the minority gap region due to the lack of initial states. Thus, at our photon energy the measured ESP around the Fermi energy should depend sensitively on additional gap states caused by bulk defects or by the surface. These states are thus the dominant contribution to the observed minority peak m1, as will be discussed in Sec.~\ref{sec:disorder} and Sec.~\ref{sec:surface}.

\subsection{Disorder effects for $\alpha=0.69$ and $\alpha=1.19$}
\label{sec:disorder}
In order to identify possible bulk disorder effects that might affect the measured ESP we have calculated the spin resolved density of states (DOS) for the nominal compositions \mrm{Co_2Mn_\alpha Si} with $\alpha=0.69$ and $\alpha=1.19$, using LSDA+CPA. Here we restrict ourselves to the DOS since bulk defects are intrinsically localized and thus are not affected by final state effects. The lattice occupations have been inferred from the formula unit model introduced by Yamamoto \etal~\cite{Yamamoto10}, assuming that the \mrm{L2_1} structure is preserved and that the formation of bulk vacancies is energetically unfavorable compared to the formation of antisites.

\begin{figure}
\centering
\subfloat[Antisite]{\includegraphics[width=\linewidth]{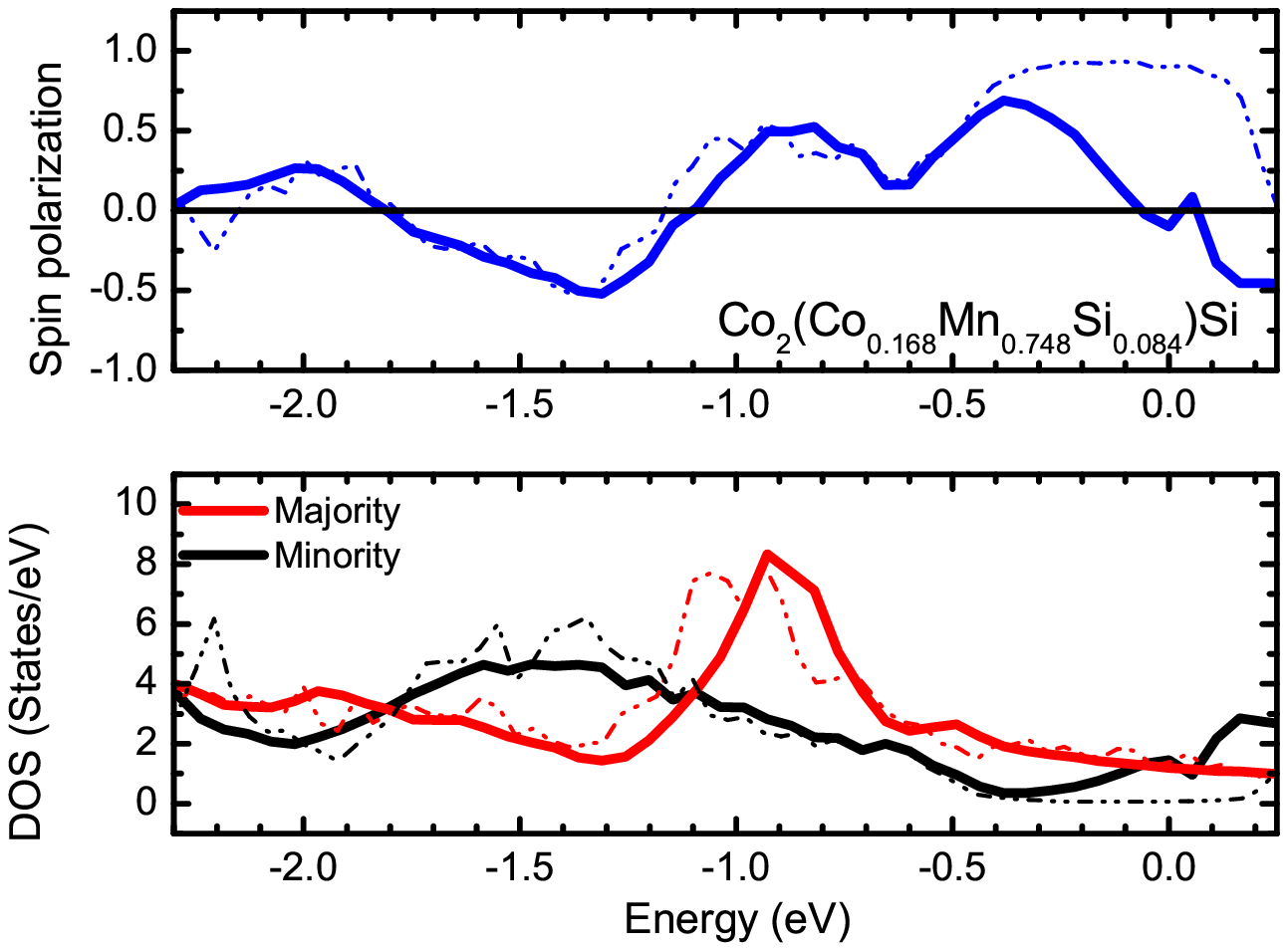}}\\
\subfloat[Vacancy]{\includegraphics[width=\linewidth]{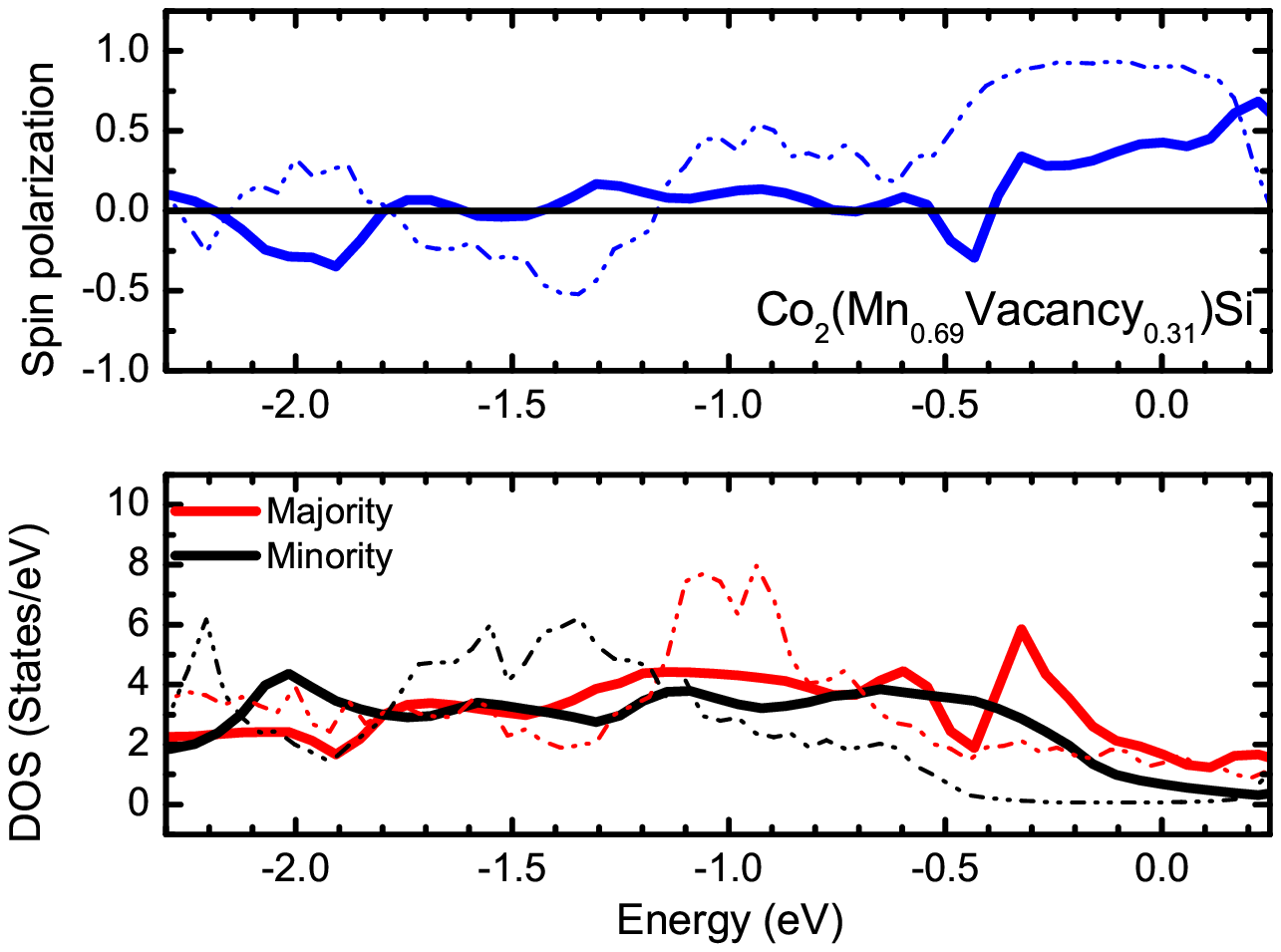}}
\caption{CPA calculations for $\alpha=0.69$. Missing Mn atoms are replaced by a) Co and Si atoms and b) vacancies. Upper panel: spin polarization. Lower panel: majority (red) and minority (black) DOS. Dashed lines represent pure LSDA calculations of stoichiometric CMS as published in Ref.~\onlinecite{Chadov09}.}
\label{fig:Mn069}
\end{figure}

This assumption has first been tested using \mrm{Co_2Mn_{0.69}Si}. Two limiting cases have been modeled: the vacancy model assumes that the positions of the missing Mn atoms are left empty; the antisite model assumes that these positions are filled with Co and Si atoms to preserve a full occupation of the \mrm{L2_1} lattice. In Fig.~\ref{fig:Mn069}a and Fig.\ref{fig:Mn069}b we show the calculated spin polarization and spin resolved DOS for the antisite model \mrm{Co_2(Co_{0.168}Mn_{0.748}Si_{0.084})Si} and the vacancy model \mrm{Co_2(Mn_{0.69}Vacancy_{0.31})Si}, respectively.  

Surprisingly, the antisite model almost reproduces all features (A\ldots E) of the measured spin polarization of CMS069. Compared to the pure LSDA calculations for stoichiometric CMS (taken from Ref~\onlinecite{Chadov09} and marked by dashed lines), the minority gap at the Fermi energy is closed in the antisite model. The additional electrons due to the excess Co are thus mainly accommodated in minority gap states. In addition, the maximum in the majority DOS appears slightly shifted towards \EF in the antisite model DOS.
 
There are two particularities of the spectra we want to highlight. First, LSDA+CPA underestimates the measured spin polarization in CMS069, predicting a Fermi level ESP of \unit{-10}{\%} instead of the measured \unit{-20}{\%}. This is astonishing since calculations tend to predict larger absolute ESP values than observed in spectra, and points to an additional surface effect that is not considered in the LSDA+CPA calculations. 
A second aspect is related to the positions of the main peaks in the minority and majority DOS. While the minority peak at -1.44\,eV fits perfectly to the peak m3 of the experimental CMS069 spectrum, the strong majority DOS peak at -0.92\,eV has no direct counterpart in the measured majority spectrum of CMS069. This points out the relevance of final state effects for the bulk related features of the spectra. Apart from that it is obvious that the formula unit model agrees well with our spectra, indicating a large contribution of bulk \mrm{Co_{Mn}} antisite defect states that contribute to the m1 feature in the minority bands of the Mn-deficient CMS069 sample.

In contrast to the antisite model, the introduction of a large degree of vacancies into the \mrm{L2_1} structure results in a massive change of the electronic structure both in the majority and the minority DOS which does not correlate with the experimental data and thus will be considered no further.

\begin{figure}
\centering
\includegraphics[width=\linewidth]{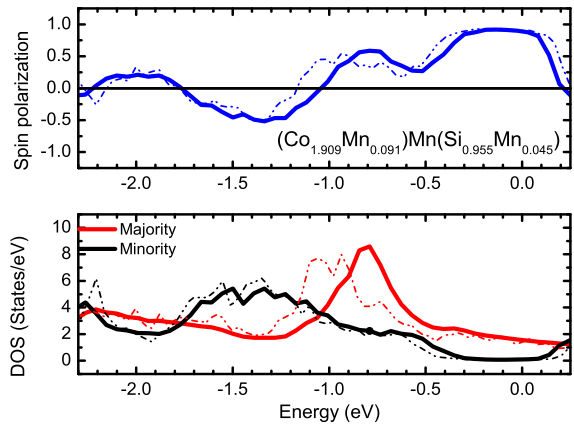}
\caption{CPA calculations for $\alpha=1.19$. Surplus Mn atoms are hosted by the Co and Si sublattice. Upper panel: spin polarization. Lower panel: majority (red) and minority (black) DOS. Dashed lines represent LSDA calculations of stoichiometric CMS}
\label{fig:Mn119}
\end{figure}

We will now focus on th Mn-rich sample CMS119. According to Ref.~\onlinecite{Yamamoto10}, increasing the \nicefrac{Mn}{Co} ratio even above the stoichiometric 1:2 should suppress the detrimental \mrm{Co_{Mn}} antisites and lead to an improved ESP. We calculated the spin polarization and spin resolved DOS for the Mn-rich \mrm{Co_2Mn_{1.19}Si} using the composition \mrm{(Co_{1.909}Mn_{0.091})Mn(Si_{0.955}Mn_{0.045})} according to the formula unit model (see Fig.~\ref{fig:Mn119}). Most importantly, the minority spin gap remains open in this configuration, though its width is slightly reduced. 

As in CMS069, the energetic positions of the ESP features (B) and (C) are almost exactly reproduced. The predicted position of (E) is the same as for CMS069, but the corresponding experimental CMS119 minority feature m3 is shifted by 0.2\,eV towards \EF (cf. Tab.~\ref{tab:Mm}). This might be a final state effect or a surface effect.
As already observed for the Mn-deficient case, the maximum of the majority DOS appears shifted to E=\unit{-0.8}{eV}, coinciding roughly with the majority feature M3 at E=\unit{-0.92}{eV} of CMS119. This might be a consequence of \mrm{Mn_{Co}} antisites which reduce the number of electrons in the majority bands. Otherwise the DOS is nearly identical to the one of stoichiometric CMS. The experimental spin polarization is qualitatively well reproduced by the simulation, but in contrast to CMS069 the calculations yield much higher absolute values of the ESP. 
We obtain a large positive spin polarization at the Fermi energy, which confirms the interpretation that \mrm{Mn_{Co}} and \mrm{Mn_{Si}} antisites do not affect the half metallic properties for $\alpha>1$ at \EF. The slight decrease in the measured ESP close to the Fermi energy for p-polarized light is not reproduced. As an explanation we propose that either the disorder induced narrowing of the minority band gap or the presence of minority surface states are responsible for the ESP reduction very close to \EF.

\subsection{Surface effects}
\label{sec:surface}
\begin{figure}
\centering
\includegraphics[width=\linewidth]{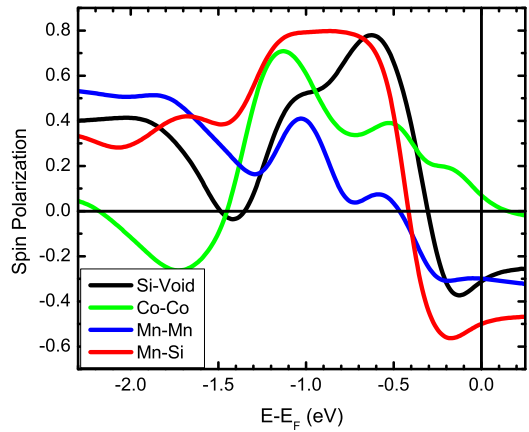}
\caption{Calculated spin polarization spectra of the CMS(100) surface, using a photon energy of 5.9~eV and s-polarized light.}
\label{fig:JM_Pol}
\end{figure}

As we have seen in the previous sections neither bulk transitions nor defects can fully explain the experimental spin resolved spectra close to \EF, and surface effects have to be taken into account.  
In order to illustrate the sensitivity of the spin polarization to the surface termination we calculated spin resolved photoemission spectra for all thermodynamically stable surface terminations of stoichiometric CMS~\cite{Hashemifar05}, using LSDA+CPA+DMFT in the framework of the one-step model of photoemission. 

\begin{figure}
\subfloat[Majority]{\includegraphics[width=0.3\linewidth,clip]{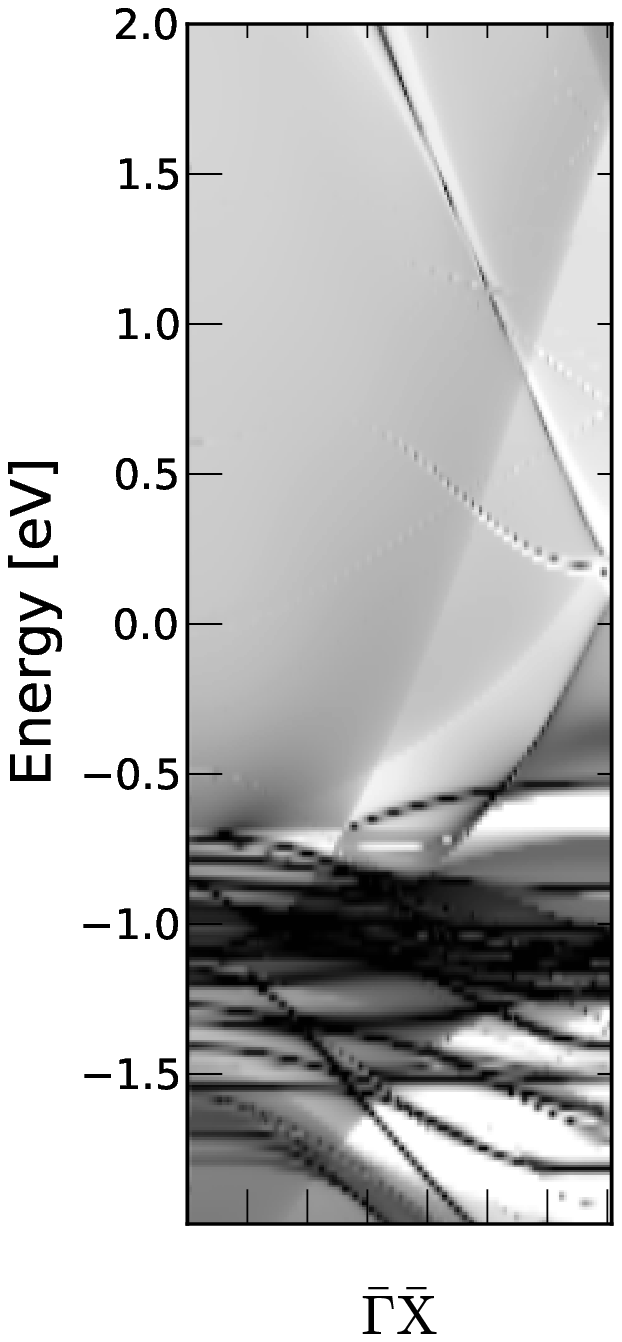}}
\subfloat[Minority]{\includegraphics[width=0.3\linewidth,clip]{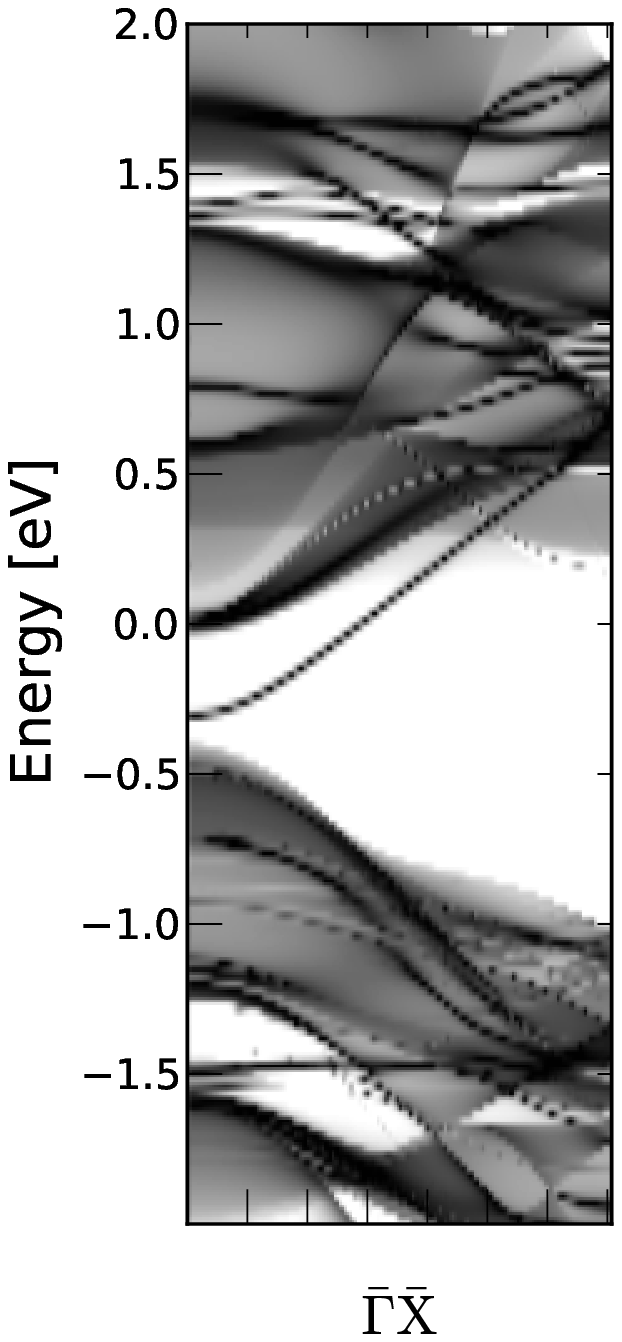}}
\caption{Surface electronic structure of the Mn-Mn terminated CMS surface along the $\bar{\Gamma}\bar{X}$ direction, calculated in the LSDA scheme.}
\label{fig:Mn_OF_Spektrum}
\end{figure}

First we concentrate on the spin polarization (Fig.~\ref{fig:JM_Pol}), which is displayed for the Co-Co, the Mn-Si, the Mn-Mn and the vacancy-Si termination. All surface terminations show a strong reduction of the spin polarization at the Fermi level, i.e.\ none of the sample surfaces is truly half-metallic in our calculation. This holds in particular for Mn-Mn terminated surface which has earlier been suggested to be half metallic~\cite{Hashemifar05}. In our calculations for that surface termination we clearly observe a dispersive minority surface state along \mrm{\Gamma X} that crosses the Fermi energy (cf. Fig.~\ref{fig:Mn_OF_Spektrum}).

For the vacancy-Si and the Mn-Si terminations corresponding to the observed LEED patters we find a good qualitative agreement with the experimental ESPs, in particular with the features (B) to (E). In CMS069, even the negative spin polarization at the Fermi energy and the bulk related minimum (E) of the spin polarization at $E-E_F=-1.4\,eV$ are well reproduced. The mono-atomic Co-Co and Mn-Mn terminations are less similar to the measured spin polarization spectra, emphasizing the strong impact of the surface termination on the photoemission spectra.  
\begin{figure}
\centering
\includegraphics[width=\linewidth]{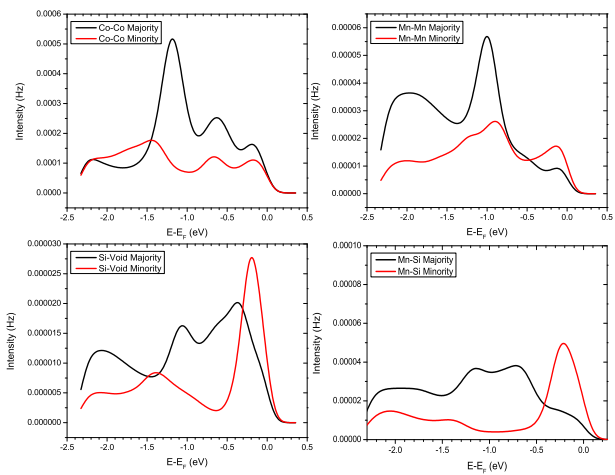}
\caption{Calculated majority and minority spectra of the CMS(100) surface, terminated by a final Co-Co, Mn-Mn, Si-void and Mn-Si layer, and using a photon energy of 5.9~eV and s-polarized light.}
\label{fig:JM_Mm}
\end{figure}

In CMS119, we find some discrepancies. First, the influence of surface states on the experimental data is less than predicted in the calculations based on stoichiometric CMS, even though CMS119 should be electronically more similar to bulk CMS than CMS069. This leads us to the conclusion that the calculations somehow overestimate the influence of surface states also in CMS069, and the negative spin polarization is the result of both bulk minority defect states and surface states. 

Second, the observed shift of the ESP minimum (E) towards \EF in CMS119 is explained neither by the bulk calculations of stoichiometric CMS nor by the CPA calculations for the non-stoichiometric systems. Only the calculated Mn-Mn surface photoemission spectra can reproduce the correct position of minimum (E). This can be seen well in the spin resolved surface spectra displayed in Fig.~\ref{fig:JM_Mm} where we observe a strong minority peak at E=\unit{-1.2}{eV}, while, in contrast to the calculations, the majority spectrum appears rather flat and unstructured. This minority feature must be dominantly surface related since it appears nowhere else in this intensity and does not show up for other surfaces. We thus conclude that a part of the CMS119 surface might be Mn-Mn terminated or at least Mn enriched. This is not in contradiction to the observed LEED patterns since, as in bulk x-ray diffraction, a fraction of B2-type ordered surface areas or point defects cannot be excluded from the surface diffraction pattern.

\section{Conclusions}
In this report we investigated the free surface of thin \mrm{Co_2Mn_{\alpha}Si}(100) films, using a combination of LEED, spin-resolved photoelectron spectroscopy and electronic structure calculations. 
Due to the little dispersion of the bulk d-bands of CMS, many features of the experimental spectra can directly be related to bulk features. However, in particular in the range of the Fermi energy where the majority bands are strongly dispersive, resonant majority interband transitions are momentum-forbidden for the selected photon energy, rendering our experiment sensitive to even small amounts of defects or surface states not only in the minority but also in the majority bands. The experimental spin resolved photoemission spectra of the free surfaces of nonstoichiometric \mrm{Co_2MnSi} films were in good agreement with LSDA+CPA calculations based on the formula unit model that assumes the formation of antisites rather than vacancies for nonstoichiometric Co2MnSi. 
Explicit photoemission calculations indicate that the surface termination also has a strong influence on the surface electronic spin polarization. Our experimental spin-resolved photoelectron spectra enable us to trace back the observed spectral features, finding bulk states at lower energies as well as surface and defect induced states close to \EF. Best agreement is obtained for the Mn-Si terminated surface wich is also supported by the LEED patterns. Polarization resolved measurements indicate majority as well as minority states with a strong \D{1}-compatible symmetry contribution at the Fermi energy, in particular in the Mn-poor compound. Here the negative spin polarization at the Fermi energy is related to minority surface and bulk antisite defect states, which is also consistent with the formula unit model. In the Mn-rich compound we find no sign of bulk defect induced minority gap states. Here the spin polarization at the Fermi energy is mainly determined by the contribution of majority electrons with \D{1} symmetry which should have a positive influence on the TMR ratio of MgO based magnetic tunnel junctions. 
\section{Acknowledgement}
Financial support through the DFG Research Unit 1464 ASPIMATT and FOR 1346, by the German ministry BMBF under contract 05KS10WMA is gratefully acknowledged. The work at Hokkaido University was partly supported by a
Grant-in-Aid for Scientific Research (A) (Grant No. 23246055) from MEXT,
Japan. We thank M.~Jourdan and C.~Herbort for providing us a
CCFA sample for the comparison of the LEED patterns.

\end{document}